\documentclass[aps,prl,twocolumn,showpacs,superscriptaddress,groupedaddress,nofootinbib]{revtex4}  
\usepackage{graphicx}  
\usepackage{dcolumn}   
\usepackage{bm}        
\usepackage{amssymb}   
\usepackage{graphicx}
\usepackage{caption}
\usepackage{mathtools}
\usepackage{subcaption}
\usepackage{breqn}

\usepackage{amssymb}
\usepackage{siunitx}
\usepackage{color}

\usepackage{perpage}
\MakePerPage{footnote}

\begin{document}


\title{Capillary filling in drop merging: dynamics of the four-phase contact point}

\author{Peyman Rostami, Günter K.~Auernhammer}
\date{\today}

\begin{abstract}
The merging of immiscible drops differs significantly from the merging of miscible drops due to the formation of a  liquid-liquid interface between drops. 
The immiscibility requires the formation of a four-phase contact point, where the drops, the gas and the substrate meet. 
We show that this point has its own unique dynamics, never studied beforehand. 
For very different scenarios, the propagation distance of this point follows scales with time like $t^{\frac 12}$. 
A model balancing the driving and dissipative forces agrees with our experiments.
\end{abstract}

\maketitle


Drop merging has an important role in industrial and natural phenomena from emulsions and microfluidics to printing technology  and metallurgy \cite{gu2011droplets,varma2016droplet, graham2017high}.   
Most of the early studies were performed with drops with identical liquids \cite{Charles:1960aa,ristenpart2006coalescence,sun2015recent}.
Later drops with different liquids are studied, most of these works focus on early stages of coalescence \cite{karpitschka2012noncoalescence, bruning2018delayed}.
Studies on dynamic wetting along liquid-liquid interfaces are rare, e.g. \cite{PhysRevFluids.5.104006}. 
In the drop merging, Marangoni tensions and the induced flow play essential roles \cite{karpitschka_riegler_2014,doi:10.1021/acs.langmuir.9b02466} and can dominate this process. 
However, one important geometric feature of two non-identical drops on a substrate seems to be overlooked so far. 
In three dimensional space $n$ phases contact in an object with $4-n$ dimensions. 
When we consider two immiscible drops on a substrate in a gas phase, $n=4$, this implies that there is a contact point in which all four phases meet. 
Here, we show that this four-phase contact point (FPCP from now on) has its own dynamics which has a characteristic time scale that differs significantly from other time scales. 

The statics of the FPCP is introduced about two decades ago \cite{MAHADEVAN:2002aa}.
So far the mechanical stability of specific configurations was analyzed \cite{Ayyad:2011aa,Zhang:2016ab}. 
Another example of capillary wetting on liquid-liquid interfaces is drop encapsulation, where the liquid with lower surface tension surrounds the other drop. 
During this process, capillary force acts in the three phase contact line and leads to encapsulate the drop \cite{koldeweij2019marangoni}. 
Despite its relevance for microfluidics \cite{Dunne:2020aa, shen2017study}, etc.~little work on drop merging of immiscible drops is done on solid surfaces \cite{li2016coalescence}. 

Capillarity driven filling in capillaries and V-shaped grooves is both long studied problem and still an active field of research. 
This is due to the fact that passive capillary flow inside macro and micro channels are used in wide range of applications like microfluidics  \cite{C2LC20799K, Seemann:2005ab}, printing technologies \cite{huang2018gravure}, capillary pumping \cite{guo2018capillary} or liquid imbibition in different scales \cite{PhysRevLett.125.127802,Seo:2018ab} . 
More than three centuries ago, Hauksbee and Jurin published first accounts on the static shape of a water meniscus in the wedge between two touching glass plates \cite{Hauksbee:1710aa,Hauksbee:1710ab,Jurin:1717ab}. 
Lucas \cite{Lucas1918}  and Washburn \cite{PhysRev.17.273} studied independently the dynamic filling of capillaries by solving the hydrodynamic problem inside the capillary, i.e., the force balance between driving and dissipating forces. 
For zero gravity condition, e.g., a horizontal capillary, only the capillary driving force and viscous dissipation force are relevant. 
By writing this force balance it can be shown that the penetration length $H$ inside the capillary scales with the square root of time Eq (\ref{Eq:Washburn}). 
The coefficient of penetration $D$ combines the radius of the capillary $r$, the cosine of contact angle $\theta$ and the ratio of surface tension $\gamma$ to viscosity of liquid $\eta$. 

\begin{equation}
H^{2}(t) = \frac{\gamma }{\eta }\frac{\cos(\theta )}{2} \,rt \equiv D^{2} \, t
\label{Eq:Washburn}
\end{equation}  

It has been shown, the structure of this solution can be used for tubes of different cross sections on different length scales \cite{C3SM51082D, WU2016195,  LIU2017234, Ouali2013, romero_yost_1996,B912235D}. 
The same scaling with time also applies for the propagation of liquid in open micro channels\cite{doi:10.1021/la9500989,doi:10.1021/acs.langmuir.6b04506,Ouali2013}. 
In these cases, the coefficient of penetration picks up  geometry parameters of the channel.
A typical example is capillary flow inside an open V-shaped channel that was studied numerically \cite{10.1115/1.1404119} and experimentally \cite{romero_yost_1996,C003728A,doi:10.1021/la9500989}.
Recently, the capillary filling inside an open micro channel which is already filled with another immiscible liquid \cite{doi:10.1021/jp2065826} was studied, which again follows the Washburn scaling $H^2 \sim t$.

\begin{figure}[tbp]
  \centering
\includegraphics[width=50mm]{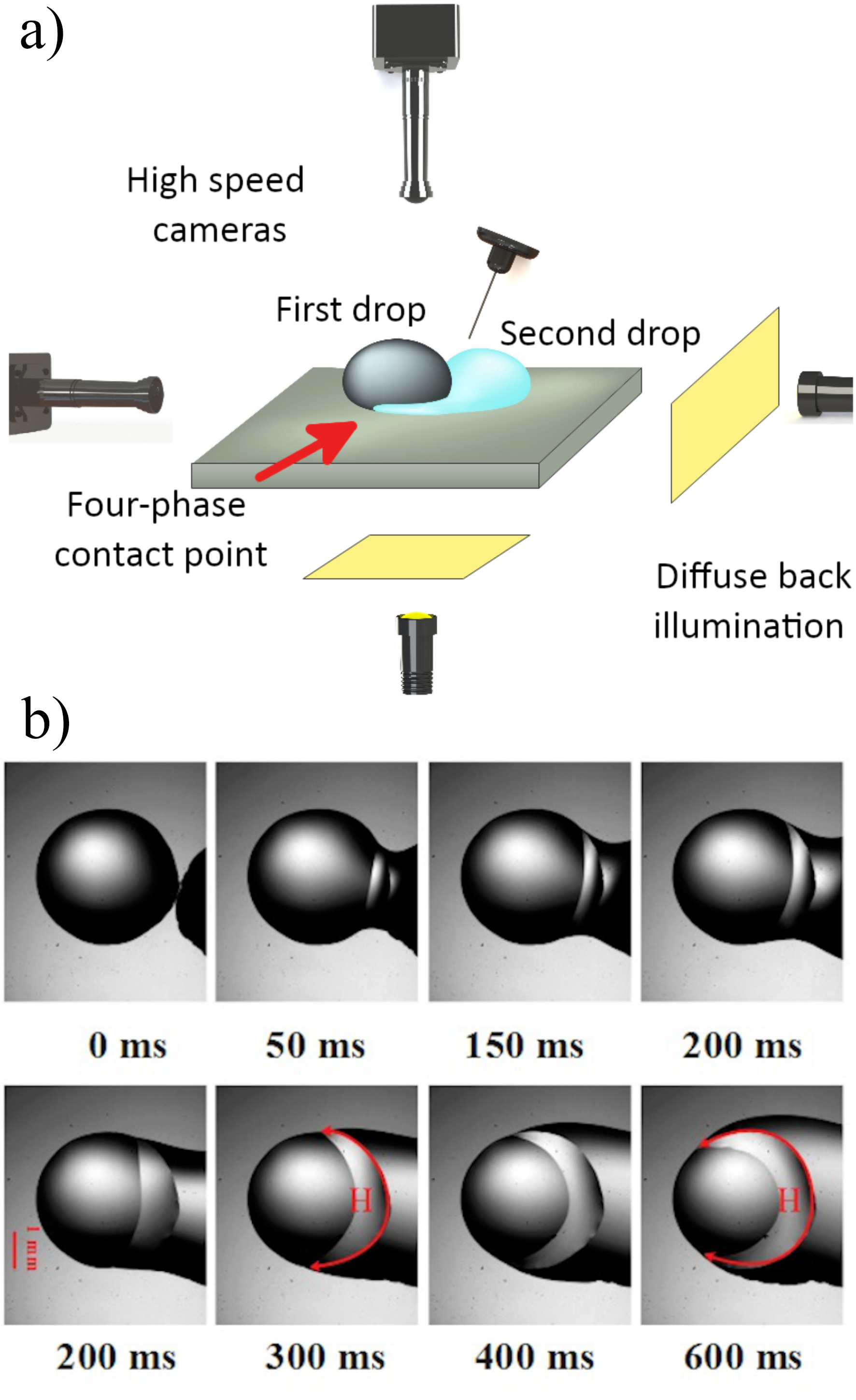}
 \caption{a) Sketch of the setup with an illustration of the FPCP. b) Time series of drop merging when the first  drop (left, here water) is already deposited and the second drop (Bromocycloheptane) comes to contact, the FPCP position is highlighted by arrows. $H$ measures the length of the contact line of the first drop that is wetted by the second drop.}\label{fig:typical_results}
\end{figure}

In this study, in this study, we go beyond these established scenarios. 
We follow the merging dynamics of  immiscible drops that initially touch along the contact line. 
We find that there is a unique dynamics of the FPCP that decouples from the rest of the merging dynamics of the drops. 
The lower surface tension of liquid engulfs the higher surface tension drop along the contact line of the drop, Fig.~\ref{fig:typical_results}(b). 
The propagation of the FPCP has strong similarities to capillary filling phenomena as described by the Lucas-Washburn equation, Eq.~\ref{Eq:Washburn},  and follows the same scaling.
To demonstrate this scaling, we systematic vary parameters: 
(i) the ratio of surface tension to viscosity of second drop, (ii) the viscosity of pre-deposited first drop and (iii) the wetting properties of the substrate.
These parameter variations allow us to develop and test a simple analytical model that describes the dynamics of the FPCP. 

We follow the drop merging with two high-speed cameras, having orthogonal directions of observation, Fig.~\ref{fig:typical_results}a). 
The drops are deposited on glass substrate with syringes pumps. 
The cameras are equipped with Navitar objectives with a maximum 12$\times$ magnification with a 2$\times$ F-mount adapter, allowing for a lateral size of the field of view of \SI{17.5}{\milli \meter}) at a frame rate of \SI{1000}{\hertz}. 
Two LED lamps with diffuser sheets provide a uniform illumination of the observation section. 
All of the experimental results are obtained under controlled environmental conditions with a relative humidity  \SI{48\pm3}{\percent} and a temperature \SI{23.0\pm0.5}{\degreeCelsius}.
Glass substrates are cleaned before performing the experiments.

We use two classes of liquids: High surface tension liquids (water, glycerine and mixtures thereof) and brominated oils as low surface tension liquids. 
At the experimental times scales analyzed in this work, the brominated oils are immiscible in the polar liquids to a good approximation. 
The purchased chemicals are used as delivered without further purification.
For particle tracking experiments, polystyrene particles are used. 
The physical properties of the used liquids are listed in Table.~\ref{table:Physical properties}.
Further details on the materials and the experimental procedure are given in the supporting information.

\begin{table}
\centering
\caption{Physical properties of liquids.}
\begin{tabular}{lrrr}
Chemicals         & \begin{tabular}[c]{@{}r@{}}Surface tension \\ (mN/m)\end{tabular} & \begin{tabular}[c]{@{}r@{}}Viscosity \\ (mPa.s)\end{tabular} & \begin{tabular}[c]{@{}r@{}}Surface\\ tension to \\ viscosity \\(m/s)\end{tabular} \\\\[-2ex]\hline\hline\\[-2ex]
Water             & 72.1 $\pm$ 0.1                                                      & 0.9321 \cite{antonpaar:aa}                                                         & 80.1                                                                                     \\
Glycrine           & 63.5 $\pm$ 0.1                                                      & 1078  \cite{calc_cheng:aa,cheng2008formula}                                                        & 0.06                                                                                  \\ \\[-2ex]\hline\\[-2ex]
Bromocyclopentane & 33.2 $\pm$ 0.1                                                      & 1.4  $\pm$ 0.05                                                        & 23.7                                                                                     \\
Bromocyclohexane  & 32.1 $\pm$ 0.1                                                      & 2.2  $\pm$ 0.05                                                          & 14.6                                                                                     \\
Bromocycloheptane & 31.5 $\pm$ 0.1                                                      & 3.9  $\pm$ 0.05                                                         & 8.1                                                                                      \\   

\end{tabular}
\label{table:Physical properties}
\end{table}

\begin{figure}[tbph]
  \centering
  \includegraphics[width=\columnwidth]{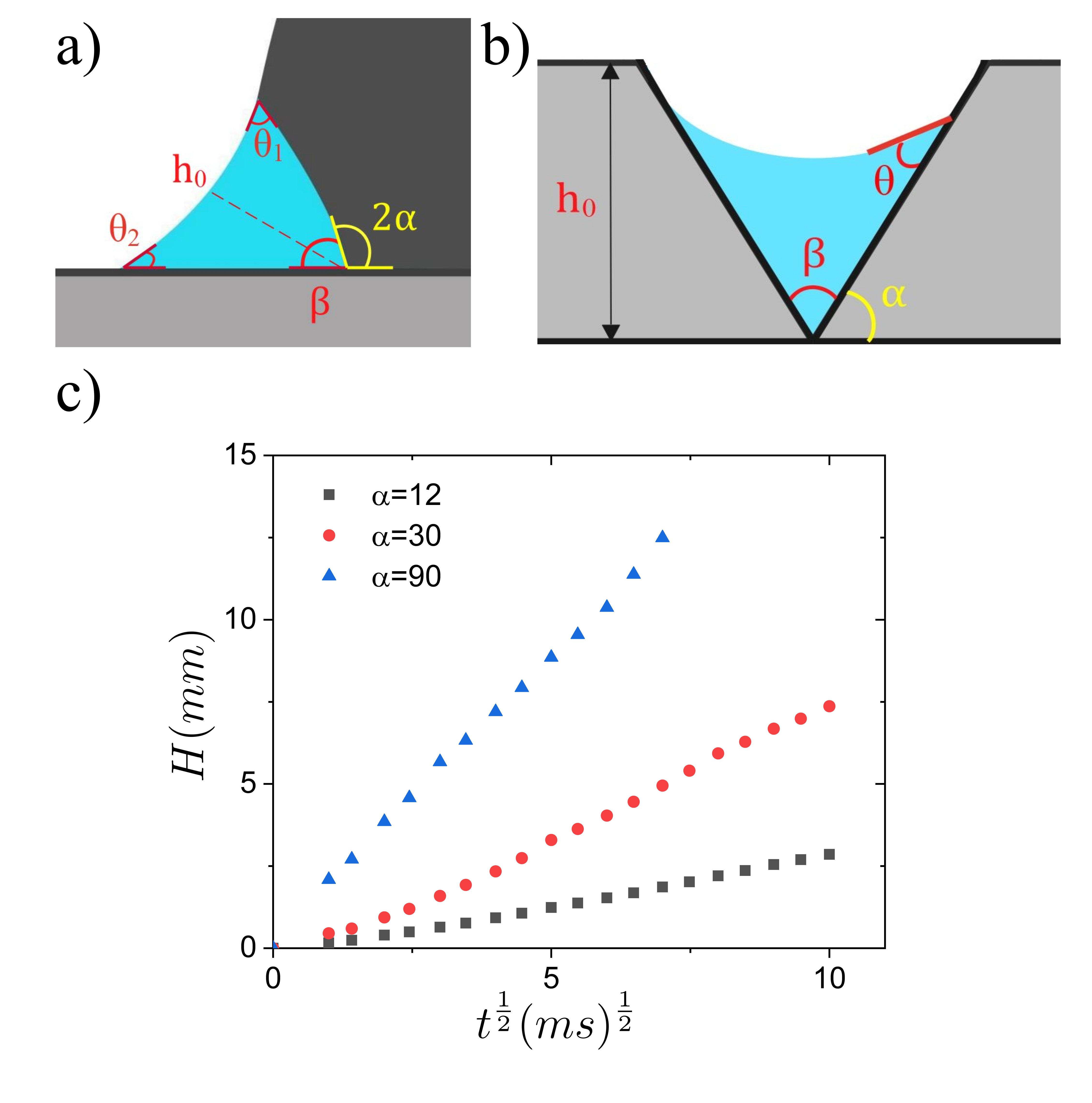}
  \caption{ a) The cross section near the FPCP illustrates the similarity to the V-groove. b) Schematic of V-groove channel: depth of channel ($h_{0}$), contact angle of the filling liquid with the walls ($\theta$) , opening angle of groove ($\beta$) and the angle between wall and horizontal line ($\alpha $). c)Typical dynamics of the FPCP. Here  for the liquid pair of water (first drop) and boromocyclohexane (second drop) on substrates with different hydrophobicity, as indicated by the contact angle of water ($\alpha$).}\label{fig:longtimeeffect}
\end{figure}

In all experiments, at contact of the two drops, two FPCPs form at both sides of the neck, where the two liquids, the substrate and the gas phase meet,  Fig.~\ref{fig:typical_results}b). 
The propagating FPCPs ''wet'' the contact line of the first drop along a length $H$, i.e., the liquid-gas-solid contact line is changed into a liquid-liquid-solid contact line. 
These FPCPs travel around the first drop and annihilate each other when contacting at the other side of the first drop.
A quantitative analysis shows that the advancing FPCP  propagates with a power law $H(t) \sim t^{\frac 12}$, Fig.~\ref{fig:longtimeeffect}c).

The cross-sectional area filled by the second drop near the FPCP is illustrated in Fig.~\ref {fig:longtimeeffect}a), along with the similarity to the filling of a V-shaped groove Fig.~\ref {fig:longtimeeffect}b).
The propagating liquid (blue)  is confined between a solid, the glass substrate, and the liquid-liquid interface to first drop. 
Despite the similarities, there is an important difference. 
The classical Washburn case considers solid walls, but in our case one boundary of the ''groove'' is the liquid of the first drop. 
This liquid-liquid interface can deform and exhibits a different hydrodynamic boundary condition for the flow of the second drop.
But still, the FPCP follows the general Washburn scaling $H(t) \sim t^{\frac 12}$. 
Can we apply the Washburn modeling to the present case? 
If yes, how much it will deviate from general phenomena because of the liquid wall and dynamic time-dependent geometry? 
To approach these questions, several key mechanisms have to be checked, including the ratio between surface tension and viscosity and the viscosity and contact angle of the first drop.

The similarity of the dynamic spreading in a groove and the classical Washburn case was shown previously \cite{romero_yost_1996,Lenormand:1984aa}. 
Romero and Yost \cite{romero_yost_1996} developed a quantitative model. 
Since this model uses one fluid and assumes solid walls, we call it the ''one fluid model''.
In this model, the coefficient of penetration  $D$ contains two contributions: i) the ratio of the surface tension and the viscosity $\gamma/\eta$ multiplied with a typical lateral dimension of the channel (here its depth $h_0$),  and ii) a geometric factor $K(\alpha ,\theta )$ that depends on the contact angle and the opening angle of the V-shaped groove. 

 \begin{equation}
H^{2}(t)=K(\alpha ,\theta )\frac{\gamma h_{0}}{\eta }t
\label{Eq:Washburn-experimental}
\end{equation}  

\begin{figure}[tbph]
  \centering
  \includegraphics[width=\columnwidth]{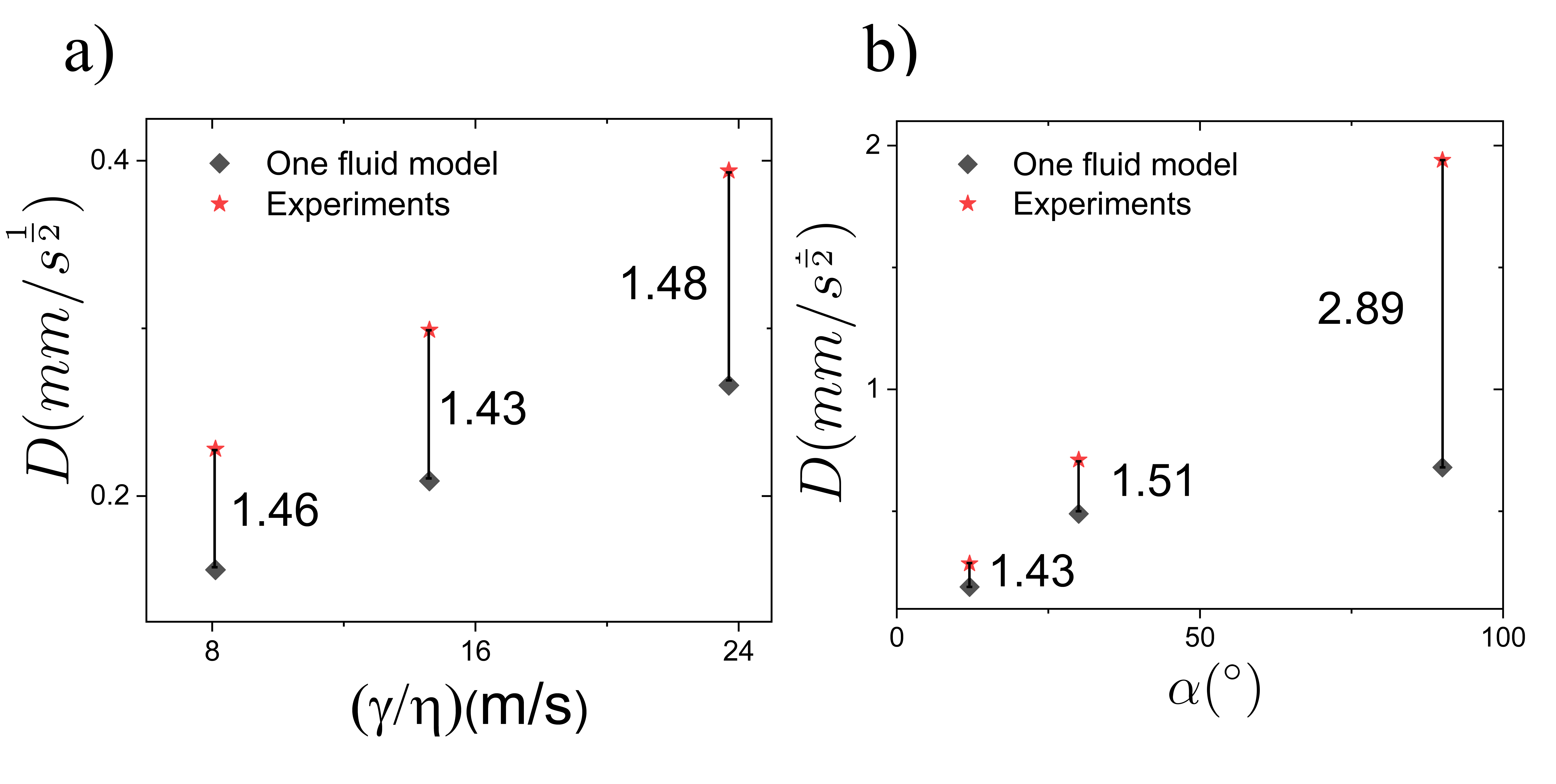}
  \caption{ a) The coefficient of penetration as measured  experimentally $D_{exp}$ and according to the one fluid model, Eq.~(\ref{Eq:Washburn-experimental}), $D_{ofm}$, using the data given in Table \ref{table:Physical properties}.
  In all cases, water is used as first drop. The numbers at the right side of the data points are the ratios $D_{exp}/D_{ofm}$. b) The coefficient of penetration as measured  experimentally $D_{exp}$ and according to the one fluid model, Eq.~(\ref{Eq:Washburn-experimental}), $D_{ofm}$  for water (first drop) and boromocyclohexane (second drop) on substrates with different hydrophobicity, as indicated by the contact angle of water ($\alpha$). }
  \label{fig:ST to viscosity}
\end{figure}

For a systematic variation of the ratio  ($\gamma/\eta$) in the spreading drop, we use liquids that are chemically very similar, with approximately the same surface tension, but different viscosities, Table \ref{table:Physical properties}. 
For all liquids combinations, the length $H$ of the wetted contact line is proportional to the square root of time.
By increasing the ratio of surface tension to viscosity, the FPCP travels faster, i.e., the coefficient of penetration is increasing, Fig.~\ref {fig:ST to viscosity}a).
There is, however, a pronounced difference between the experimental data and the predictions of the one fluid model, even when taking the drop height of the first drop as an upper bound of the groove depth $h_0$.

A similar systematic difference is observed when changing the opening angle of groove $\beta$, Fig.~\ref {fig:ST to viscosity}b).
We use surface modifications to change the contact angles.
Obviously, this changes the contact angles of both drops. 
Thus, a surface modification changes the opening angle of the groove $\beta$ and the drop height  Fig.~\ref {fig:longtimeeffect}a). 
The one fluid model Eq.~(\ref {Eq:Washburn-experimental}) predicts that the dynamics of the FPCP speeds up when the opening angle of groove ($\beta$) decreases, i.e., for higher contact angles of the first drop.
Experimentally, the FPCP propagates significantly faster for more hydrophobic surfaces. 

The coefficient of penetration shows qualitatively in the right trend on  $\gamma/\eta$ and the opening angle of the groove. 
However, the one fluid model systematically underestimates the dynamics of the FPCP, Fig.~\ref{fig:ST to viscosity}b). 
Potential reasons for this difference are:
i) The capillary driving originates from two different surfaces, the liquid-solid interface and the liquid-liquid interface. 
The unbalanced capillary force that drives spreading differs on these two interfaces. 
Additionally, the contact between the two liquid induces Marangoni tensions that can induce flow in the first drop \cite{doi:10.1021/acs.langmuir.9b02466}.
ii) The geometry changes during the spreading of the second drop, because the first drop deforms during the spreading process. 
iii)  The hydrodynamic boundary condition at the liquid-liquid interface differs from the one on the substrate and depends on the viscosity ratio of the two liquids.
iv) The contact angles depend on contact line velocity which might scale with the velocity of FPCP \cite{Snoeijer:2013aa}. 
All theses effects only influence the prefactor $K$ but don't change the basic scaling with time $H \sim t^{\frac 12}$. 
The capillary driving and the viscous dissipation are localized in a narrow region close to the moving contact lines. 
Consequently, a small region close to the FPCP should dominant for its dynamics. 

How does the dynamics of the FPCP relate to other characteristic time scales in drop merging? 
The inertial time  scale $T_{in}$ is the square root of the ratio between mass of displaced liquid ($M$) and capillary ($\gamma$). 
Taking  the displaced liquid volume as two cones with a length $\frac{H}{2}$ and base radius of $\frac{R}{2}$ at $\varphi =5^{\circ}$, Fig.~S3, we get  $M\sim \rho H R_{2}^{2}$, with the density of second liquid  $\rho$.
For calculating the characteristic time of inertial regime, we use the crossover time defined by  \cite{Biance:2004aa} ($T_{in}\sim (\frac{\rho \gamma H}{\eta ^{2}})^{\frac{1}{8}}\sqrt{\frac{\rho H R_{2}^{2}}{\gamma }}$).  
The viscous dissipation time scale $T_{vis}$ is the ratio between the mass of displaced liquid and viscous force $\eta H$:  $T_{vis}\sim {\rho H R_{2}^{2}/H\eta  }$. 
If we estimate the momentum diffusion time, the same characteristic time scale can be calculated.
Finally, from the modeling of the dynamics of the FPCP a time scale arises, which is the ratio between viscous and the capillary force,  $T_{fpcp}\sim {H\eta/\gamma}$.  
By putting the physical values into the mentioned time scales will leads to $T_{fpcp}\sim 1ms$, $T_{in}\sim 100ms$ and $T_{vis}\sim 200ms$.
The characteristic time of the FPCP is much shorter than the other one. 

The general case of filling of groove was solved by Yost and coworkers \cite{romero_yost_1996}.
In this solution the authors argue that the scaling of $H \sim \sqrt t$ is rather general and applies also for the tip of the wetting liquid in a partially filled V-shaped groove. 
They show that the final result is the same as given in Eq.~(\ref{Eq:Washburn-experimental}). 
The geometrical differences between horizontal tubes and grooves can be covered by a prefactor to the scaling. 
Due to this equivalence and for sake of simplicity, we adopt the approach of \cite{doi:10.1021/la9500989}, knowing the scaling also applies for partially filled grooves. 
The tip of the wetting liquid as calculated by Yost and coworkers \cite{romero_yost_1996,doi:10.1021/la9500989} corresponds to the FPCP in our case. 

\begin{figure}[tb]
\centering
\begin{minipage}{.5\textwidth}
  \centering
   \includegraphics[width=\columnwidth]{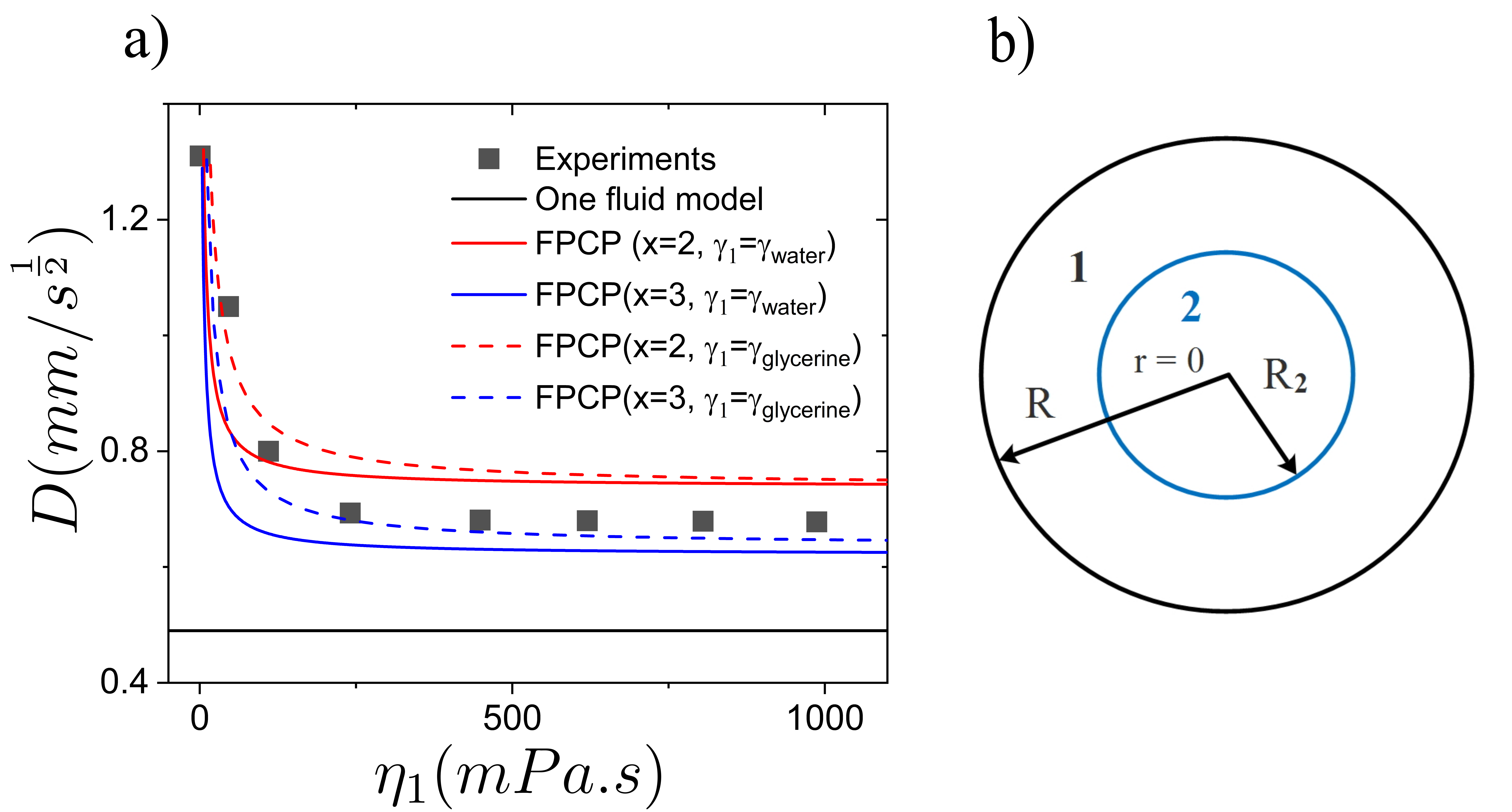}
\end{minipage}
  \caption{a) Coefficient of penetration as the function of first drop viscosity with boromocyclohexane as second drop. The experimental results are compared with one fluid model and four phase contact point model with different $x$ values.  b) Schematic for the modeling of the flow with two fluids inside the cylinder, the inner region is assumed as filling liquid or second drop and outer section is the first drop ($R_{2}$ and $R$ are the radii of inner and outer region).} 
  \label{fig:viscosity effect}
\end{figure}

To capture the  basic differences mentioned above, for the liquid wall, we consider a multiphase flow with a geometry of concentric cylinders, where the boundary between the inner and outer cylinder is assumed radially undeformable, Fig.~\ref{fig:viscosity effect}b). 
This outer cylinder mimics the first drop (region 1), the inner cylinder the second drop (region 2). 
The dynamics of capillary flow inside a tube can be written by the force balance between driving mechanisms and dissipation forces \cite{doi:10.1021/la9500989}.
We give details of the calculation in the SI. 
 \begin{equation}
H(t)=\sqrt{\frac{h_{0}K(\gamma, \alpha, \theta_1, \theta_2 )}{4\pi \eta _{\mathrm{eff}}}t}
\label{Eq:Height of capillary}
\end{equation}
Here, we introduced an effective viscosity $\eta_{\mathrm{eff}}$, a modified geometric factor $K(\gamma, \alpha, \theta_1, \theta_2 )$, Eq.~(S10),  and the ratio of the outer and inner radius $x = R/R_2 \ge 1$. 
 \begin{equation}
\eta _{\mathrm{eff}}=\frac{1}{2}(\eta _{2}+\frac{\eta _{1}\eta _{2}}{\eta _{1}+2\eta _{2}(x^{2}-1)})
\label{Eq:Effective viscosity}
\end{equation}
This model depends only weakly on the value of $x$. $2 \le x \le 4$  are compatible with our data, see also Fig.~S2. 
However, the results remain in the same range for higher value of $x$, we consider two cases of $x=2$ and $x=3$.


Increasing the viscosity of the pre-deposited first drop allows for a simple test of the crossover between a liquid (low viscous) and a quasi solid (highly viscous) wall. 
Since the additives for changing the viscosity typically also change the surface tensions, the driving force is indirectly a weak function of the viscosity. 
This weak dependence has, however no significant influence on the quality of the model comparison, compare Fig.~\ref{fig:viscosity effect}a). 
Increasing the viscosity of the pre-deposited  first drop also increases $\eta _{\mathrm{eff}}$. 
Consequently, the viscous force  $F_{\eta}$ increases and the filing rate decreases, see Eq.~(\ref{Eq:Height of capillary}). 

To illustrate this, we compare the merging of bromocyclohexane drop with different pre-deposited drops of water, glycerine, and mixtures thereof, Fig.~\ref{fig:viscosity effect}. 
The apparent contact angle of all the first drops stays in the range ${25}^{\circ}\le \theta \le{30}^{\circ}$. 
The surface tension differences of the liquids pairs are also close, $\SI{33}{\milli \newton\per \meter} \le  \gamma _{1}-\gamma _{2} \le \SI{28}{\milli \newton\per \meter}$.
Consequently, the driving forces varies by less than \SI{20}{\percent}. 
In contrast, the coefficient of penetration reduces by \SI{50}{\percent}. 

Experimentally, the coefficient of penetration only depends on the viscosity of the first drop for small enough viscosities, Fig.~\ref{fig:viscosity effect}. 
For viscosity ratio $\eta_1/\eta_2 \gtrsim 250$, the coefficient of penetration remains constant.
Increasing the viscosity is almost equivalent to the shift from a liquid-liquid interface at one of the walls and to liquid-quasi solid interface.
The one fluid model does not cover this viscosity dependence, Fig.~\ref{fig:viscosity effect}. 
In contrast our model above gives a good description of the behavior. 
Despite the simplifications in our model and the omitted dynamic effects, the difference between our model and the experimental data is below 10\% for $D$.

Our model of the dynamics of the FPCP is in almost quantitative agreement with the experimental data. 
The remaining differences between our model and experiments can be explained by several facts.
Probably other forces play a role that were neglected in the present model, i.e., Marangoni stress between the two liquids.
Additionally, the FPCP is highly dynamic and the geometry of the wedge changes during  the experiments.
Obviously, neglecting these effects does not seem to influence the dynamics of the FPCP significantly. 

In this contribution, we have investigated the dynamics of the FPCP (four-phase contact point)  where two liquids, the substrate and the gas phase are in contact. 
During drop merging this FPCP shows a dynamics that is independent of other processes like contact line velocity. 
The dynamics of this point follows the same scaling as the capillary flow inside the tube as originally described by Lucas and Washburn.
The motion of the FPCP is consistent with $H(t)\approx D t^{\frac{1}{2}}$ for all investigated cases, with the coefficient of penetration $D$.
The geometry of the moving FPCP resembles the spreading of liquid in a V-shape groove, of which the dynamics can be reduced to the Lucas-Washburn case.
Different to the classical case of the  V-shape grooves, in the present case one of the ''walls'' of the groove is liquid, i.e., the pre-deposited first drop. 
Despite this deformable boundary and the other differences  to the classical case, the governing mechanisms are the same.
Various material and geometry parameter can be included in a simple model that describes the dynamics of the FPCP almost quantitatively. 

The FPCP which is introduced here, not only reveals a unique dynamic but also leads to higher mobility of the liquid.
Since we substitute one solid wall with a liquid wall higher mobility is expected and it can lead to higher efficiency of systems.
This new phenomena should be relevant in microfluidics, patterned coating and printing technology. 

\section*{Acknowledgement}
Funded by the German Research Foundation (DFG) within the Collaborative Research Centre 1194 “Interaction between Transport and Wetting Processes”, Project-ID 265191195, subprojects A02 (P.R), A06 (G.K.A).
In addition, the authors wish to thank Benedikt Straub and Stefan Michel for fruitful discussions.

\bibliography{literature} 
\bibliographystyle{rsc} 

\end{document}



\title{Supplementary Materials: \\
Capillary filling in drop merging: dynamics of the four-phase contact point}

\author{Peyman Rostami, Günter K.~Auernhammer}
\date{\today}
\maketitle
\setcounter{equation}{0}
\setcounter{figure}{0}
\setcounter{table}{0}
\setcounter{page}{1}
\makeatletter
\renewcommand{\theequation}{S\arabic{equation}}
\renewcommand{\thefigure}{S\arabic{figure}}
\renewcommand{\bibnumfmt}[1]{[S#1]}
\renewcommand{\citenumfont}[1]{S#1}

\section{Materials and Methods}
\subsection{Materials}
We used the following equipments for experiments (see Fig.~1a of the main text),  two high-speed cameras (Photron miniAX200 and Photron minAX100), two LED lamps (SCHOTT KL 2500 LED) and syringe pump ( Legato® 185 Syringe Pump, kd Scientific co.). 
Glass substrates from Thermo Scientific Menzel-Gläser, Microscope Slides, 76$\times$ 26 $\times$ 1 (mm) are used.
The cleaning process of the glass substrates is done by immersing them subsequently in tetrahydrofuran (Acros Organics Co. 99.6\%), acetone (Fisher Scientific Co.  95\%) and isopropanol (Fisher Scientific Co.~95\%). 
In each liquid, ultrasound is applied using a VWR Ultrasonic Cleaner Co. USC-TH for \SI{15}{\minute}). 
Ultra pure water (${0.055}(\mu S.cm^{-1})$)  was prepared by a MicroPure UV/UF, Thermo Scientific Co.).
The purchased chemicals bromocyclopentane (Sigma-Aldric Co. 98\%), bromocyclohexane (Sigma-Aldric Co. 98\%), bromocycloheptane (Sigma-Aldric Co. 98\%),  glycrine (Sigma-Aldric Co.) are used as delivered without further purification.
For particle tracking experiments, Polystyrene particles (PS/Q-R-L2580 micro particles GmbH) are used. 

\subsection*{Experimental procedure}

Droplets of the two different liquids are deposited subsequently on the substrate using a steel needle connected to the syringe pump with a constant volume flow rate.
The volume flow rate, typically a few $\mu L.s^{-1}$, depends on the experimental conditions.
The experiments are done in a manner that perimeter of first drop was kept the same, i.e., the volume of the first drop was adapted depending on its contact angle. 
Only when the first drop has relaxed completely, the second drop is deposited, next to the first one, such that the two drops come into contact with each other, during spreading of the second drop.
Then, drop merging occurs. 
 Systematically, the liquid of the second drop has the lower surface tension and spreads around the pre-deposited first drop along the three phase contact line. 
However, second liquid does not spread over the entire surface of the first drop. 

\section{Derivation of the four-phase contact point model}

\begin{figure}[tbph]
  \centering
  \includegraphics[width=50mm , scale=0.4]{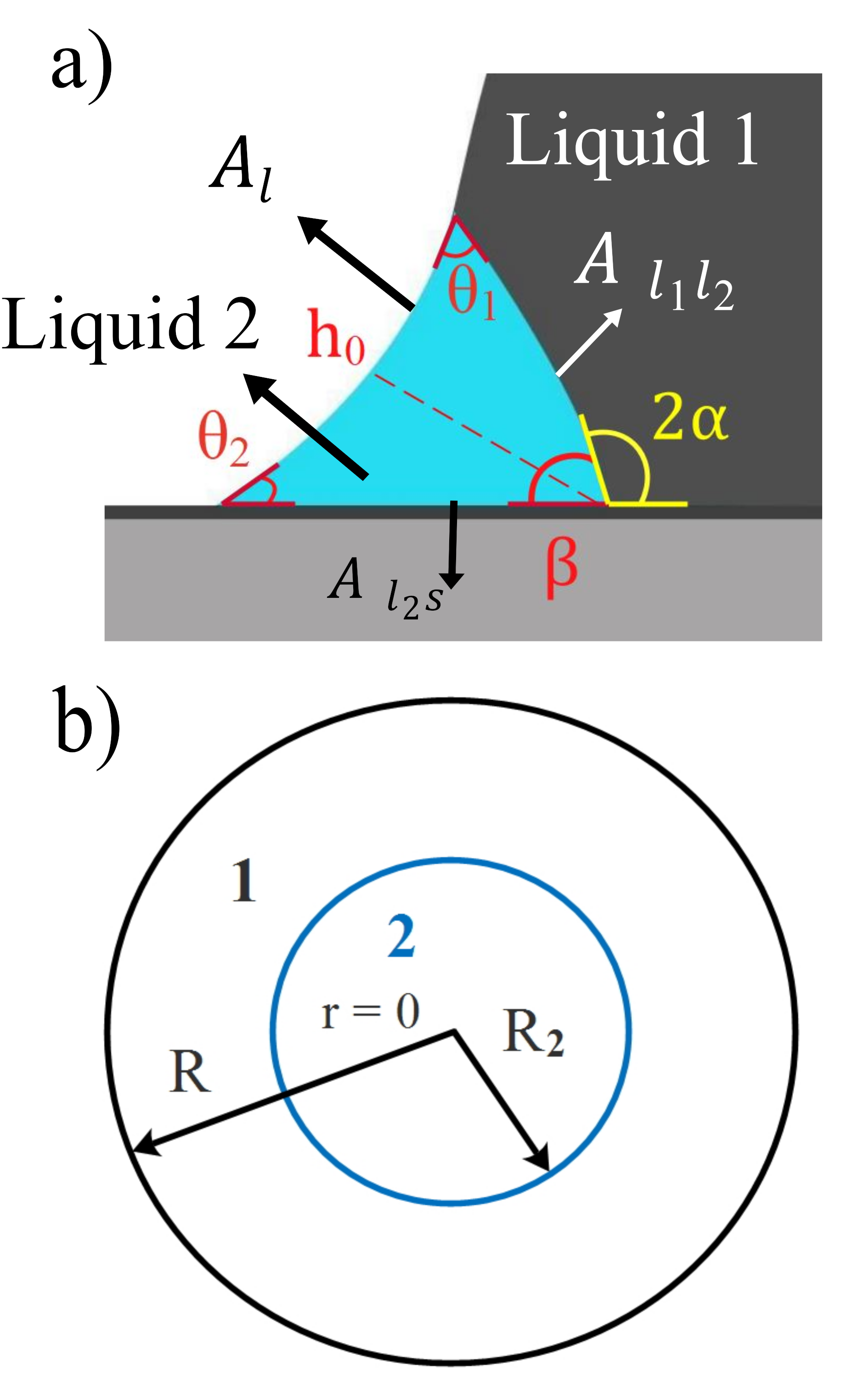}
  \caption{ a)Cross section through the V-shaped groove formed by the first drop (liquid 1), the substrate and the second drop (liquid 2). $2\alpha$, $\beta$, $\theta _{1}$ and $\theta _{2}$ are contact angle of the first drop in contact to the second drop, the opening angle of the equivalent groove, the contact angle of second drop with first drop and the contact angle of second drop and substrate, respectively. $A_{l }$, $A_{l_{1}l_{2}}$ and $A_{l_{2}S}$ are interfacial areas between the second liquid and the gas phase, between the two liquids and  between the second liquid and the substrate, respectively. $h_{0}$ is the depth of equivalent groove. b) Schematic of two fluids inside the cylinder, the inner region is assumed as filling liquid or second drop and outer section is the first drop ($R_{2}$ and $R$ are the radii of inner and outer region). }
  \label{fig:SITwo phase flow}
\end{figure}

In the merging of non-identical and immiscible drops on a surface, one important geometric feature seems to be overlooked so far. 
This geometric feature calls four-phase contact point (FPCP), where the two liquids, the substrate and the gas phase meet. 
To describe the dynamics of the FPCP, we use its similarity to the capillary filling of V-shaped grooves. 
The dynamics of capillary flow inside a tube can be written by the force balance between driving mechanisms and dissipation forces \cite{doi:10.1021/la9500989}.
After simplification for tubes, the final form consists of inertial term on the left side and forces due to surface tension,  $F_\gamma$ and viscous dissipation, $F_\eta$ on right side Eq.~(\ref{Eq:Force balance}).
 \begin{equation}
m(H) \frac{\mathrm{d^2} H}{\mathrm{d} t^2}= F_{\gamma }- F_{\eta }
\label{Eq:Force balance}
\end{equation}  
The left hand side of Eq.~(\ref{Eq:Force balance}) can be neglected for small tubes diameters. 
This approximation also holds in our case, because the weight of the liquid close to the tip of four-phase contact point is small.
Then at the end the force balance between viscous dissipation and capillary force determine the dynamic of four-phase contact point.

As mentioned in the main part of the manuscript, the general case of filling of groove was solved by Yost and coworkers \cite{romero_yost_1996}.
In this solution the authors argue that the scaling of $H \sim \sqrt t$ is rather general and applies also for the tip of the wetting liquid in a partially filled V-shaped groove. 
The geometrical differences between horizontal tubes and grooves can be covered by a prefactor to the scaling. 
Due to this equivalence and for sake of simplicity, we adopt the approach of \cite{doi:10.1021/la9500989}, knowing the scaling also applies for partially filled grooves. 
The tip of the wetting liquid as calculated by Yost and coworkers \cite{romero_yost_1996,doi:10.1021/la9500989} corresponds to the FPCP in our case. 

To capture the basic difference to our experiments, the liquid wall, we consider a multiphase flow with a geometry of concentric cylinders, where the boundary between the inner and outer cylinder is assumed radially undeformable, Fig.~\ref{fig:SITwo phase flow}b). 
This outer cylinder mimics the first drop (region 1), the inner cylinder the second drop (region 2). 
The dynamics of capillary flow inside a tube can be written by the force balance between driving mechanisms and dissipation forces \cite{doi:10.1021/la9500989}.

\subsection{Viscous force}
First, we evaluate the viscous dissipation force. 
We take a pragmatic approach an average the viscous dissipation on the solid wall (similar to \cite{doi:10.1021/la9500989}) and the viscous dissipation at the liquid ''wall''. 
The part between the second drop and solid wall is simple and the viscosity would be equal to the viscosity of second liquid. 
To calculate the dissipation at the liquid-liquid interface, we consider the geometry given in Fig.~\ref{fig:SITwo phase flow}b) with a radially undeformable boundary between the two liquid regiions. 
At the  liquid-liquid boundary between region 1 and 2, the tangental stress acting on the boundaries must be equal:
 \begin{equation}
\tau_{2}(R_2)=\tau_{1}(R_2)=- \frac{\Delta p}{L}
\label{Eq:tau1}
\end{equation}
In cylindrical coordinates, this stress can be written (with $u$ being the velocity parallel to the cylinder axis): 
 \begin{equation}
\tau (r)=\frac{\eta }{r}\frac{\partial }{\partial r}(r\frac{\partial u(r) }{\partial r})
\label{Eq:tau}
\end{equation}
By putting  Eq.~(\ref{Eq:tau}) into  Eq.~(\ref{Eq:tau1}) and integrating twice and applying the boundary conditions the velocity for second regions can be calculated:

 \begin{equation}
u_{2}=-\frac{G}{4\eta _{2}}r^{2}+\frac{G}{4}(\frac{R^{2}-{R_2}^{2}}{\eta _{1}}+\frac{{R_2}^{2}}{\eta _{2}})
\label{Eq:final u2}
\end{equation}
Here, we used the abbreviation $G=\frac{\Delta P}{L}$. 
The flow resembles the flow inside a cylinder of two co-flowing liquids, i.e., we can estimate the effect of the outer liquid (in region 1) on the flow of the inner liquid (in region 2).

In this simple model, region 2 mimics the second drop which fills the groove. 
For the force balance we also need to calculate the pressure difference ($\Delta P$): 
 \begin{equation}
\Delta P=\frac{4H u_2^{\mathrm{avg}}}{(\frac{{R_2}^{2}}{2\eta _{2}}+\frac{R^{2}-{R_2}^{2}}{\eta _{1}})}
\label{Eq:Two phase H-P equation}
\end{equation}
 \begin{equation}
F_{\eta_{2} }=\pi {R_2}^{2} \Delta P
\label{Eq:Viscous force}
\end{equation}
Here, we introduced an effective viscosity $\eta_{\mathrm{eff}}$, Eq.~(\ref{Eq:Effective viscosity}) and  the ratio of the outer and inner radius $x = R/R_2 \ge 1$. 
 \begin{equation}
\eta _{\mathrm{eff}}=\frac{1}{2}(\eta _{2}+\frac{\eta _{1}\eta _{2}}{\eta _{1}+2\eta _{2}(x^{2}-1)})
\label{Eq:Effective viscosity}
\end{equation}
Setting $x=1$ corresponds to the one fluid model, i. e., for $x = 1$ the Hagen-Poisseuille results are obtained, with $\eta_{\mathrm{eff}}=\eta_{2}$.

\section*{Capillary driving force}

Now, we turn to calculating the capillary driving force.  
To do so,  the flow inside the groove in our experiment is considered, Fig.~\ref{fig:SITwo phase flow}b).
As in the classical derivation \cite{doi:10.1021/la9500989}, we will combine the above calculation of the flow in a circular tube with the driving force in the experimental situation. 
For a complete adaption, We now derive the prefactor $K$ in our experimental geometry, Fig.~\ref{fig:SITwo phase flow}a). 
We assume that the contact angle in liquid-solid side ($\theta_{2}$) as well as liquid-liquid interface ($\theta_{1}$) are independent of the speed of the  four-phase contact.
In general, these two contact angles have different values.
With these assumptions the total spreading free energy can be written as in Eq.~(\ref{Eq:Free energy equation}).
To calculate the free energy, the method  developed by Good \cite{good1973rate} has been used. 
Again, Eq.~(\ref{Eq:Free energy equation}) additionally contains the liquid-liquid interface, which is new in this approach.
 \begin{equation}
E= (\gamma _{sl_{2}}-\gamma _{s})A_{sl_{2}}+(\gamma _{l_{2}l_{1}}-\gamma _{l_{1}})A_{l_{1}l_{2}}+\gamma _{l_{2}}A_{l_{2}}
\label{Eq:Free energy equation}
\end{equation}
 Here  $\gamma _{sl_{2}}$, $\gamma _{s}$,$\gamma _{l_{1} l_{2}}$, $\gamma _{l_{2}}$ and $\gamma _{l_{1}}$ are the interface tensions between solid-liquid2, solid-gas, liquid1-liquid2, liquid2-gas, liquid1-gas, respectively. 
$A_{l_{2}s}$, $A_{l_{1}l_{2}}$ and $A_{l_{2}}$ are the contact areas of the liquid2-solid, liquid1-liquid2 and liquid2-gas.

With same analogy which we used for one fluid model \cite{doi:10.1021/la9500989}, and having Young condition for $\theta_{2}$ and Neuman condition for $\theta_{1}$, one can calculate the capilary driving force as  Eq.~(\ref{Eq:Capillary driving force}).
 \begin{equation}
F_{\gamma }=-\frac{dE}{dH} =h_{0} \, K(\gamma, \alpha, \theta_1, \theta_2 )
\label{Eq:Capillary driving force}
\end{equation}
For a compact notation, the $K$ function is introduced.
\begin{dmath}
K(\gamma, \alpha, \theta_1, \theta_2 )=\frac{1}{\sin(\frac{\alpha }{2})}  [  \gamma _{l_{2}}\cos(\theta _{2}) + \gamma _{l_{1}}-\gamma _{l_{1}l_{2}} - 
\gamma _{l_{2}}(\frac{\cos(\frac{\alpha }{2} )(\frac{\alpha }{2} -\theta _{1})}{\sin(\frac{\alpha }{2} -\theta _{1})}+\frac{\cos(\frac{\alpha }{2} )(\frac{\alpha }{2} -\theta _{2})}{\sin(\frac{\alpha }{2} -\theta _{2})})]
\label{Eq:K}
\end{dmath}

Putting Eqs.~(\ref{Eq:Viscous force}) and (\ref{Eq:Capillary driving force}) into Eq.~(\ref{Eq:Force balance})  leads to the modified version of the Washburn equation for the spreading of the four-phase contact point:

 \begin{equation}
8\pi \eta _{\mathrm{eff}}H\frac{\mathrm{d} H}{\mathrm{d} t}=K(\gamma, \alpha, \theta_1, \theta_2 )h_{0}
\label{Eq:Force balance }
\end{equation}
 \begin{equation}
H(t)=\sqrt{\frac{h_{0}K(\gamma, \alpha, \theta_1, \theta_2 )}{4\pi \eta _{\mathrm{eff}}}t}
\label{Eq:Height of capillary}
\end{equation}

\subsection*{Time dependence of capillary filling}

\begin{figure}[tb]
\centering
\begin{minipage}{.5\textwidth}
\centering
\includegraphics[width=\columnwidth]{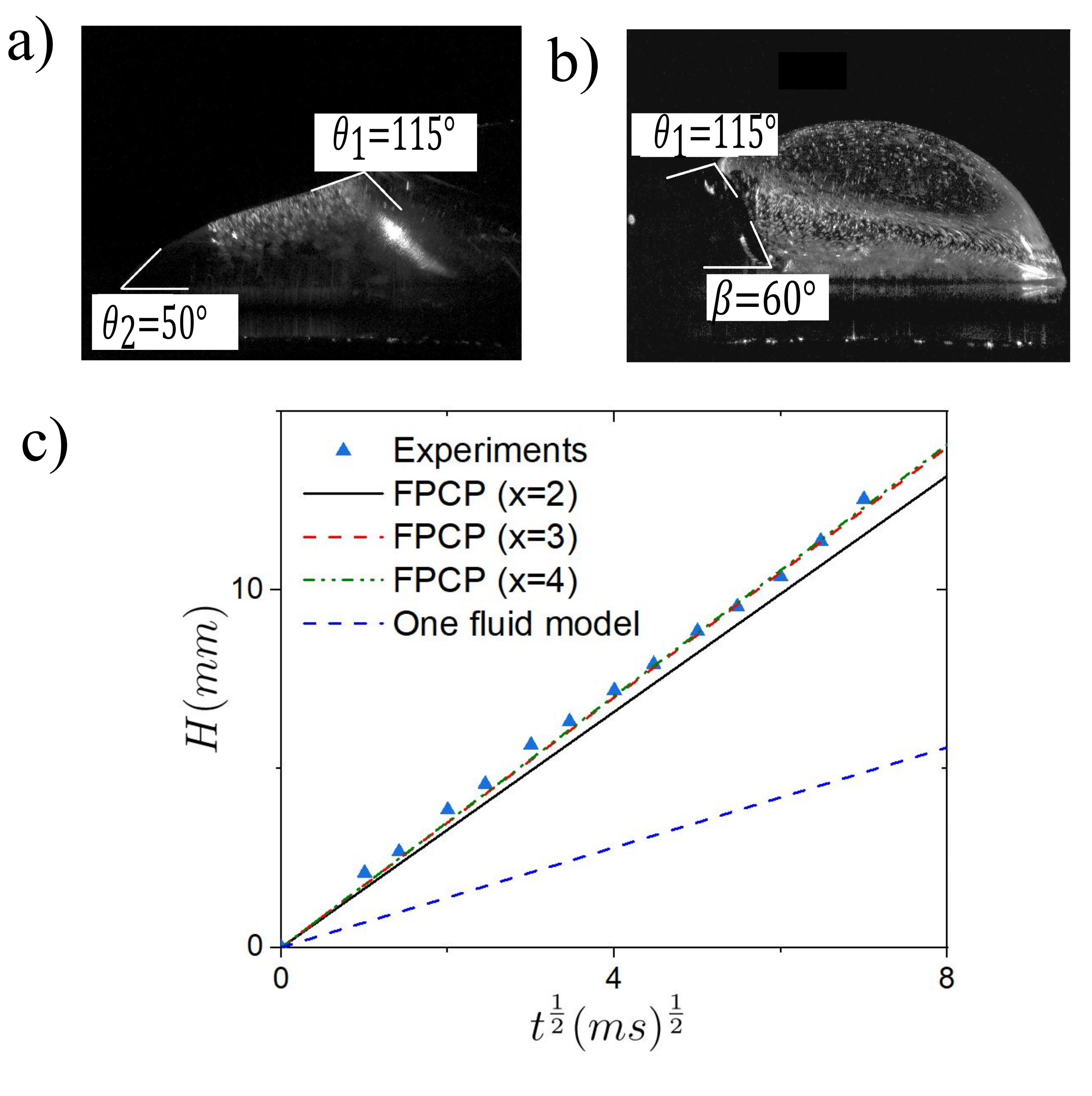}
\end{minipage}
\caption{ a) Final stage of drop merging when the bromocyclohexane contains tracer particles ($\theta_{1}$ and $\theta_{2}$ is measured). b) The final stage of drop merging for water and bromocyclohexane which the water drop contains tracer particles. $\theta_{1}$ and $\beta$ are measured. c) Comparison between experimental results of meniscus height for water and boromocyclohexane on hydrophobic substrate with the presented model Eq.~(\ref{Eq:Height of capillary}) and one fluid model.} 
\label{fig:Two phase flow}
\end{figure}

To compare theoretical models to our experimental data, we determine the opening angle of the groove, the contact angles and the depth of the groove $h_{0}$. 
To measure all contact angles precisely, we add tracer particles subsequently to each of the liquids, and repeat the experiments twice on the same surface. 
In each repetition of the experiment only one liquid was labeled with tracer particles, see Fig.~\ref{fig:Two phase flow}a)  and b). 
From these images, taken after all hydrodynamic flows have died out, we measure the contact angle of the spreading liquid on the substrate $\theta_{2} \approx \SI{50\pm 2}{\degree}$, the contact angle at liquid-liquid interface $\theta_{1} \approx \SI{115\pm 2}{\degree}$, the opening angle of the wedge $\beta \approx \SI{60\pm 2}{\degree}$, and the depth of the groove is $h_1 \approx \SI{1.6\pm 0.05}{\milli \meter}$ in this case. 

The best overlap between the theoretical model is given for a ratio of the radii of $x = 3$. 
We note, however, that, due to the simplicity of our model, the ratio of radii $x$ might slightly depend on the neglected parameters like Marangoni tensions. 
The calculated effective viscosity with with $ x = 2 $ is $\eta_{\mathrm{eff}}= \SI{1.18}{\milli\pascal\second}$.
When calculating $K(\gamma,\alpha,\theta_1,\theta_2)$, special care has to be taken not to overestimate the values of the surface tensions. 
The surface tension of first liquid $\gamma_{l_{1}}$ should be the surface tension of first liquid slightly saturated with the second liquid, i.e., water saturated with boromocyclohexane has \SI{65\pm 1} {\milli\newton\per\meter}.
The interfacial tension between the two liquids $\gamma_{l_{1}l_{2}}$ is \SI{26\pm 1} {\milli\newton\per\meter}.
Using Eq.~(\ref{Eq:Height of capillary}), the coefficient of penetration is $D \approx \SI{1.6}{\milli\meter\per \second^{\frac{1}{2}}}$.
When $x=3$ is used, the coefficient of penetration is $D \approx \SI{1.7}{\milli\meter\per \second^{\frac{1}{2}}}$.
The comparison to the one fluid model,  shows that our presented model is a significant improvement in describing the dynamics of the FPCP (four-phase contact point), Fig.~\ref{fig:Two phase flow}c).

\section*{Contact line velocity}
\begin{figure}[tbp]
  \centering
  \includegraphics[width=0.8\columnwidth]{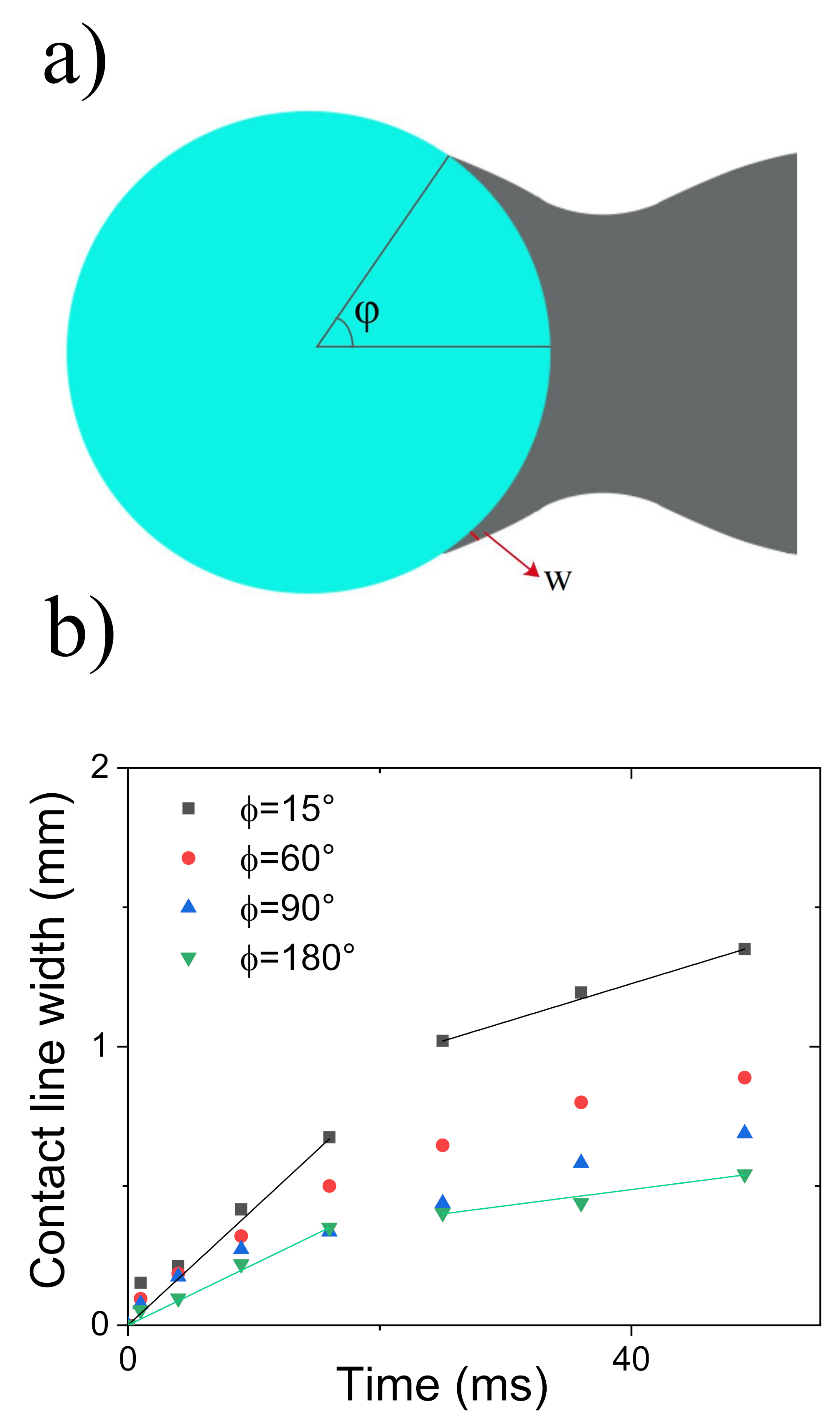}
  \caption{ a) Schematic of two drops merging from top, $\phi$ shows the geometry which we used to measure the contact line velocity and $w$ is the measured value as width of the liquid tongue. b) $w$ as a function of time, showing different regimes of spreading.}
  \label{fig:contactlinevelocity}
\end{figure}

Is there a  difference between the dynamics of the four-phase contact point and that of the contact line? 
To be able to comment on this point, we measured the width of the liquid tongue behind the four-phase contact point ($w$),(Fig.~\ref{fig:contactlinevelocity}).
We determined the contact line velocity perpendicular to the contact line of first droplet.
To do so, we  define a few parameters.
The first parameter is the position which we measured contact line velocity.
For this parameter we consider a polar system of coordinates in which ($\phi$) is the angle between first contact point of two drops and the considered position.
The second parameter would be the width of liquid tongue ($w$)  over time (Fig.~\ref{fig:contactlinevelocity}). 
We observe two regimes in contact line velocity. 
None of these regimes follows the time dependence of the four-phase contact point. In contrast, this is similar to the case of the spreading of a liquid drop  \cite{Biance:2004aa}, because of effect of inertia in very early time steps.
We compare the contact line velocity (perpendicular to the contact line) with the four-phase contact point velocity. 
The maximum velocity of contact line  (at $\phi= {15}^{\circ}$) is 6 times slower than the four-phase contact point velocity. 
The different scaling with time and this strong difference in velocity support our statement that the FPCP has its own dynamics that is independent of the  other processes in the merging of non-identical drops.

\bibliography{literature} 
\bibliographystyle{rsc} 